\documentclass[aps,12pt,tightenlines]{revtex4}
\usepackage{graphicx}

\begin{document}

\title{Constraining superheavy dark matter model of UHECR with SUGAR data}
\author{Hang Bae Kim$^a$} \email{HangBae.Kim@ipt.unil.ch}
\author{Peter Tinyakov$^{a,b}$} \email{P.Tinyakov@cern.ch}
\affiliation{$^a$Institute of Theoretical Physics,
University of Lausanne, CH-1015 Lausanne, Switzerland\\
$^b$Institute for Nuclear Research, 60th October Anniversary
prospect 7a, 117312, Moscow, Russia}
\begin{abstract}
Models which associate the origin of the ultra-high energy cosmic rays
(UHECR) with decays of relic superheavy particles predict the
anisotropy of UHECR flux toward the Galactic center.  We use the
existing SUGAR data, which covered the Galactic center best so far, to
look for such a signal and limit the fraction of UHECR produced by
this mechanism. The absence of anisotropy toward the Galactic center
in the SUGAR data implies at 95\% confidence level that the fraction
of SHDM-related cosmic rays should be less than 20\% at
$E>1\times10^{19}$~eV and less than 50\% at $E>4\times10^{19}$~eV.
\end{abstract}
\maketitle

\section{Introduction}

The origin of ultra-high energy cosmic rays (UHECR) is a long-standing
puzzle. Observations of cosmic rays with energy above $10^{19}$ eV
reveal that the GZK cutoff \cite{gzk} in the energy spectrum is not
observed by AGASA \cite{agasa} (though it seems to be consistent with HiRes
\cite{hires}), and that the arrival directions of cosmic rays are quite
isotropic at large scales but cluster at small angles\cite{cluster}.
This is a perplexing situation for seeking UHECR sources. The
isotropic distribution of UHECR arrival directions, in the absence of
strong magnetic field which can produce it, either Galactic or
extraGalactic, implies that the sources are distributed over a few
hundred Mpc for which the matter distribution looks homogeneous,
while the absence of the GZK cutoff implies that the UHECR sources are
within the GZK radius which is about 50 Mpc. The observed correlations
of UHECR with BL Lacertae objects \cite{bllacs} do not make the situation
easier: although perfectly consistent with clustering, BL Lacs as
sources of UHECR cannot explain super-GZK events under assumption of
standard physics \cite{Kachelriess:2003yy}.  

A possible way to reconcile these contradictory facts is to assume
that, apart from UHECR accelerated in BL Lacs, there exists a second,
GZK-violating component \cite{Berezinsky:2002vt} originating from
decays or annihilation of superheavy particles within the GZK radius
\cite{shdm,bdk02}. These superheavy particles are produced in the
early universe \cite{kt99} and constitute a fraction of cold dark
matter. The latter is concentrated in galactic halos. 
Since we are 8.5 kpc away from the Galactic center, the clear
signature of superheavy particle decays is the
significant excess of the UHECR flux from the direction of the
Galactic center \cite{dt98}. This kind of anisotropy toward the
Galactic center (or toward any other strong source) has not been
observed by the Northern hemisphere arrays
\cite{Hayashida:1999ab,bsw99}. However, Northern arrays do not see the
Galactic center directly, and the small statistics of cosmic rays with
energy above the GZK cutoff makes it difficult to draw strong
constraints.

In this paper we use the SUGAR data \cite{sugar} to examine the viability
of the superheavy dark matter (SHDM) scenario of UHECR, for SUGAR
is the unique experiment which operated in the Southern hemisphere
before Pierre Auger Observatory and covered the Galactic center.
We use the Kolmogorov-Smirnov (K-S) test to check the
consistency of the SUGAR data with the arrival direction distribution
expected from superheavy particle decays in the Galactic halo.  Our
purpose is to limit a fraction of UHECR which are produced by this
mechanism.

We perform the analysis in two steps. In Sects.~II and III we analyze
the SUGAR data and derive bounds on the relative weight $\eta$ of the
{\em halo component} in the total UHECR flux. These bounds are
presented in Figures~3 and 4. They depend exclusively on the halo
profile and statistics of SUGAR data. We then move in Sect.~IV to the
implications of these constraints for the superheavy dark matter
models. The difference with the previous case is that SHDM
contribution contains {\em both Galactic and extragalactic}
parts. Their relative weight depends on a particular SHDM model
(spectrum and maximum energy) and on absolute dark matter density in
the halo of our Galaxy. Because of the uncertainties in these parameters 
we separate this question from the more robust analysis of sects.II and III.
We will see, however, that in SHDM models the extragalactic contribution
is generically subdominant in the energy range of interest,
and therefore boudns of Secs.II and III apply to SHDM part of cosmic rays
as a whole.
We present our conclusions in Sect.V. 

\section{Galactic halo models and SUGAR data}

We assume that superheavy particles constitute a part of cold dark
matter and their spatial distribution follows the dark matter
density.  We examine two different Galactic halo models.
The first one is the isothermal (ISO) halo model \cite{iso},
which is characterized by the density profile
\begin{equation}
\label{eq-iso}
\rho_{\rm ISO}(R) \propto \frac{1}{R^2+R_c^2}.
\end{equation}
The second one is the Navarro-Frenk-White (NFW) model \cite{nfw96},
which is obtained from the $N$-body simulation and thus can be more realistic
\begin{equation}
\label{eq-nfw}
\rho_{\rm NFW}(R) \propto \frac{1}{R(R+R_s)^2}.
\end{equation}
Here $R$ is the distance from the Galactic center,
$R_c$ is the scale of the halo core,
and $R_s$ is the characteristic scale of the NFW model.
For our galaxy, the size of $R_c$ is supposed to be a few kpc.
The value of $R_s$ is expected to be $10$--$20$ kpc 
on the basis of correlation between the predicted total mass and the
value of $R_s$ \cite{nfw96}. In our analysis, we leave $R_c$ and
$R_s$ as free parameters and use $R_c=4$ kpc and $R_s=15$ kpc as
prototype values.

For cosmic rays coming from the Galactic halo, the expansion of the
universe and the evolution of halo can be neglected.  We also neglect
the effects of the Galactic magnetic field on the propagation of UHECR
and will make a comment on it afterward.  Since we consider the halo
model which is spherically symmetric about the Galactic center and the
distribution of sources follows the halo profile, the cosmic ray flux
can be calculated as
\begin{equation}
\label{eq-flux}
f(\theta) = \frac{1}{4\pi} \int_{0}^{r_{\rm max}(\theta)}
	L(\sqrt{r^2-2rR_0\cos\theta+R_0^2})\;dr,
\end{equation}
where $L(R)$ is the luminosity of cosmic ray sources, $\theta$ is the
angle measured from the Galactic center direction, which is related to
the Galactic coordinates $l$ and $b$ by $\cos\theta=\cos b\cos l$,
$r_{\rm max}(\theta)=R_0\cos\theta+\sqrt{R_{\rm
max}^2-R_0^2\sin^2\theta}$, $R_{\rm max}$ is the extension of the
Galactic halo, and $R_0\sim 8.5$~kpc is the distance to the Earth from
the Galactic center.  In the case of decay, the luminosity $L(R)$ is
proportional to the number density of superheavy particles, that is
$L(R)\propto n_{\rm SH}(R)\propto\rho_{\rm Halo}(R)$.  In the case of
annihilation, $L$ is proportional to $n_{\rm SH}^2$.  Since the
density profile has a peak at the Galactic center, the annihilation
case predicts stronger anisotropy than the decay case and leads to
stronger constraints on the halo contribution to UHECRs.  In the
following, we will consider the decay case only.

To obtain the arrival direction distribution predicted from the
Galactic halo model, we use the Monte-Carlo simulation based on the
flux (\ref{eq-flux}).  In simulating cosmic ray data detected by an
array of detectors, we also need to consider the exposure function of
the detector array.  Here, we use the simple geometrical exposure
function which is a function of declination $\delta$ only
\begin{equation}
\label{exposure}
h(\delta) = \frac{1}{\pi} \left[
\sin\alpha_m\cos\lambda\cos\delta+\alpha_m\sin\lambda\sin\delta \right]
\end{equation}
where $\lambda$ is the latitude of the detector array,
$\theta_m$ is the zenith angle cut,
\begin{equation}
\xi = \frac{\cos\theta_m-\sin\lambda\sin\delta}{\cos\lambda\cos\delta},
\end{equation}
and $\alpha_m=0$ for $\xi>1$, $\alpha_m=\pi$ for $\xi<-1$ and
$\alpha_m=\cos^{-1}\xi$ otherwise.

The SUGAR array, located in the Southern hemisphere, has the latitude
$\lambda=-30.53^\circ$.  It covered the Galactic center direction best
so far, because other detectors are all located in Northern hemisphere
and cover the Galactic center marginally at best.  Therefore, the
SUGAR data is most suited at the moment for the study of cosmic rays
coming from the Galactic halo, though in the near future Pierre Auger
array will provide much better data. The published SUGAR data contain
events with zenith angle up to 72$^{\circ}$ \cite{sugar}.  We use the
data with zenith angle smaller than 60$^\circ$ considering increasing
uncertainty at large zenith angles. Figure~1 shows arrival directions
of 316 SUGAR events with zenith angles $z\le60^\circ$ and energy
$E\ge1.0\times10^{19}$~eV (upper panel), and simulated data obtained
using the flux (\ref{eq-flux}) for two halo models (lower panels).
Note that the NFW model with
$R_s=15$ kpc and the isothermal model with $R_c=4$ kpc predict similar
concentration of events around the Galactic center.

%
%
\begin{figure}
\begin{center}
\includegraphics[width=0.7\textwidth]{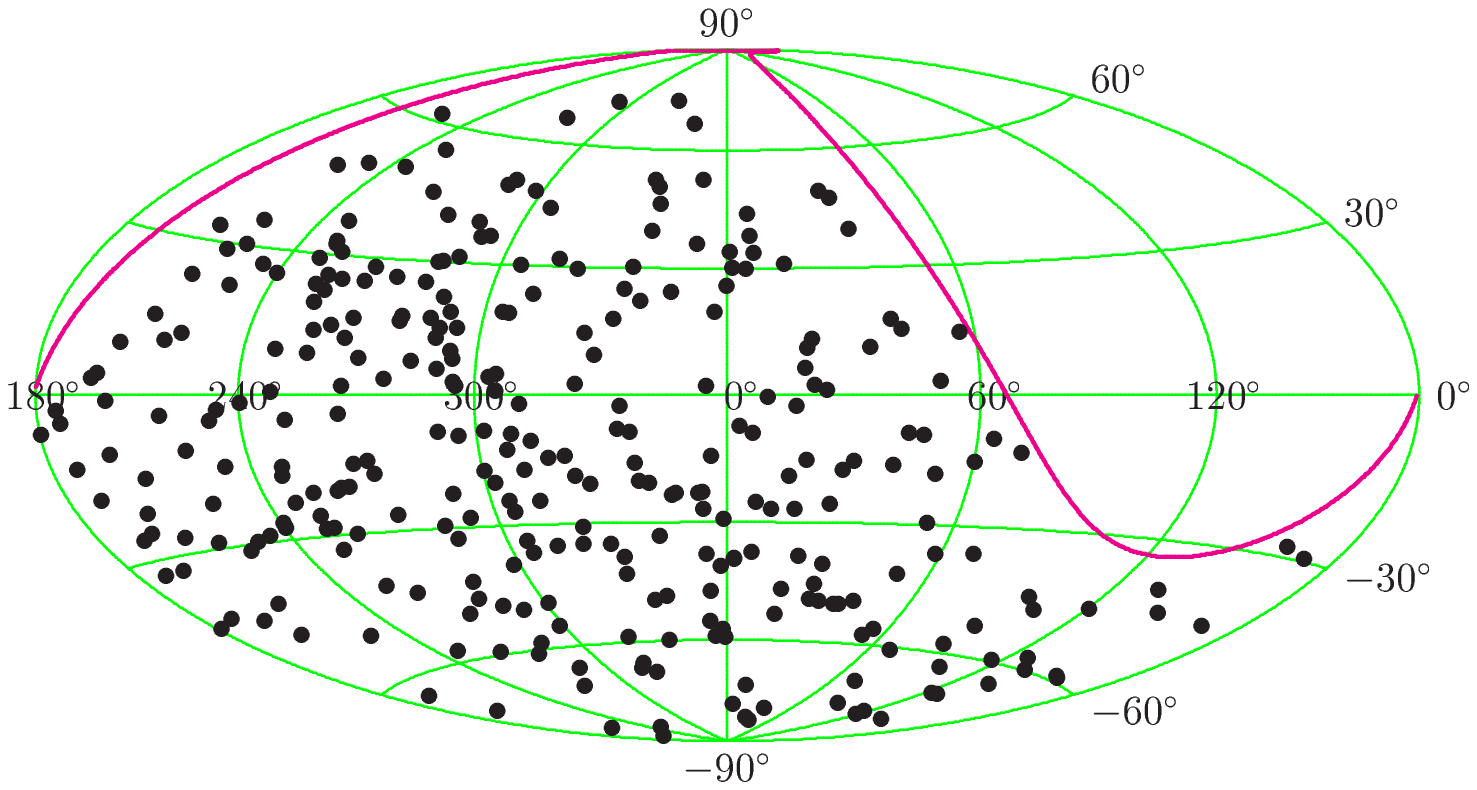}
\includegraphics[width=0.7\textwidth]{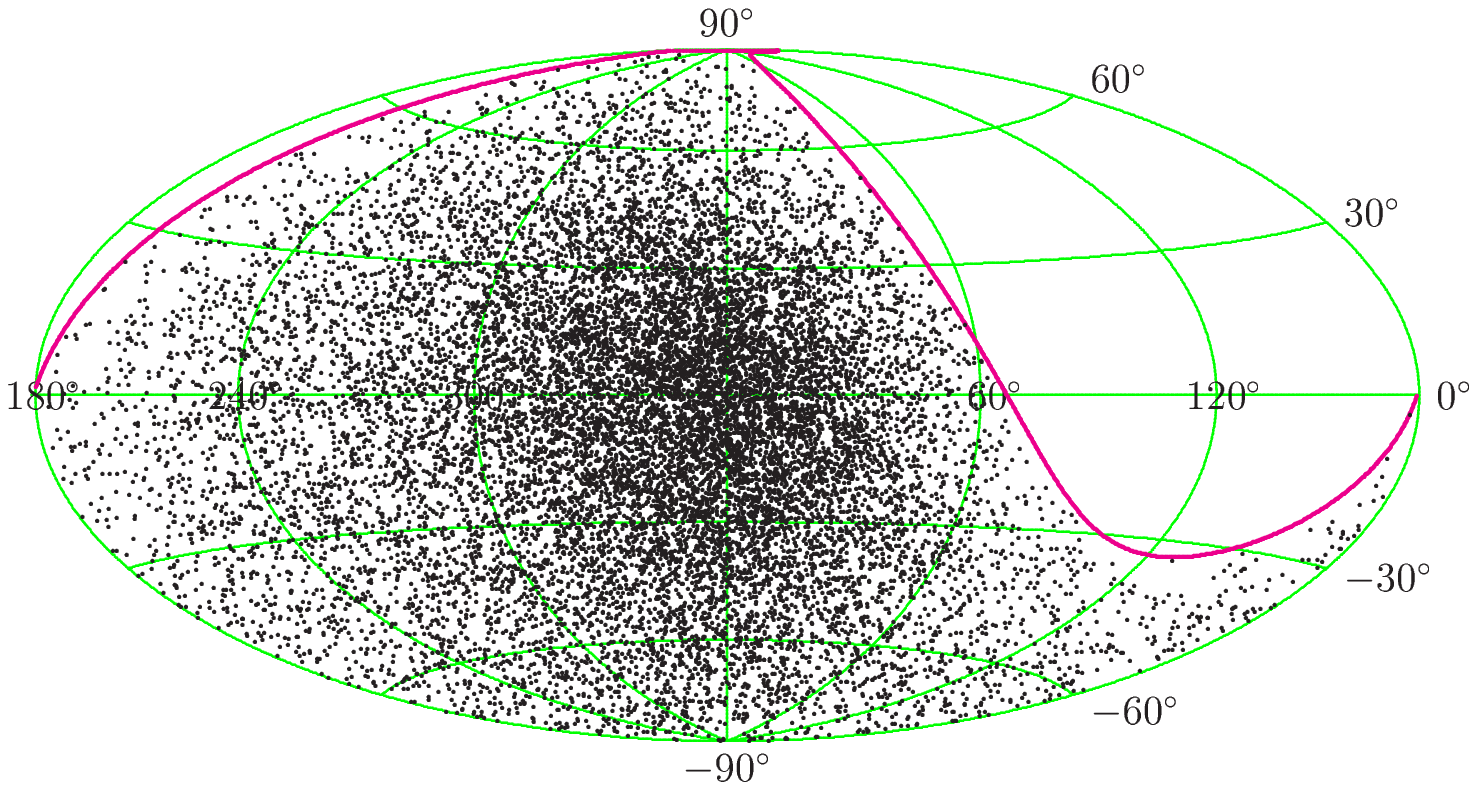}
\includegraphics[width=0.7\textwidth]{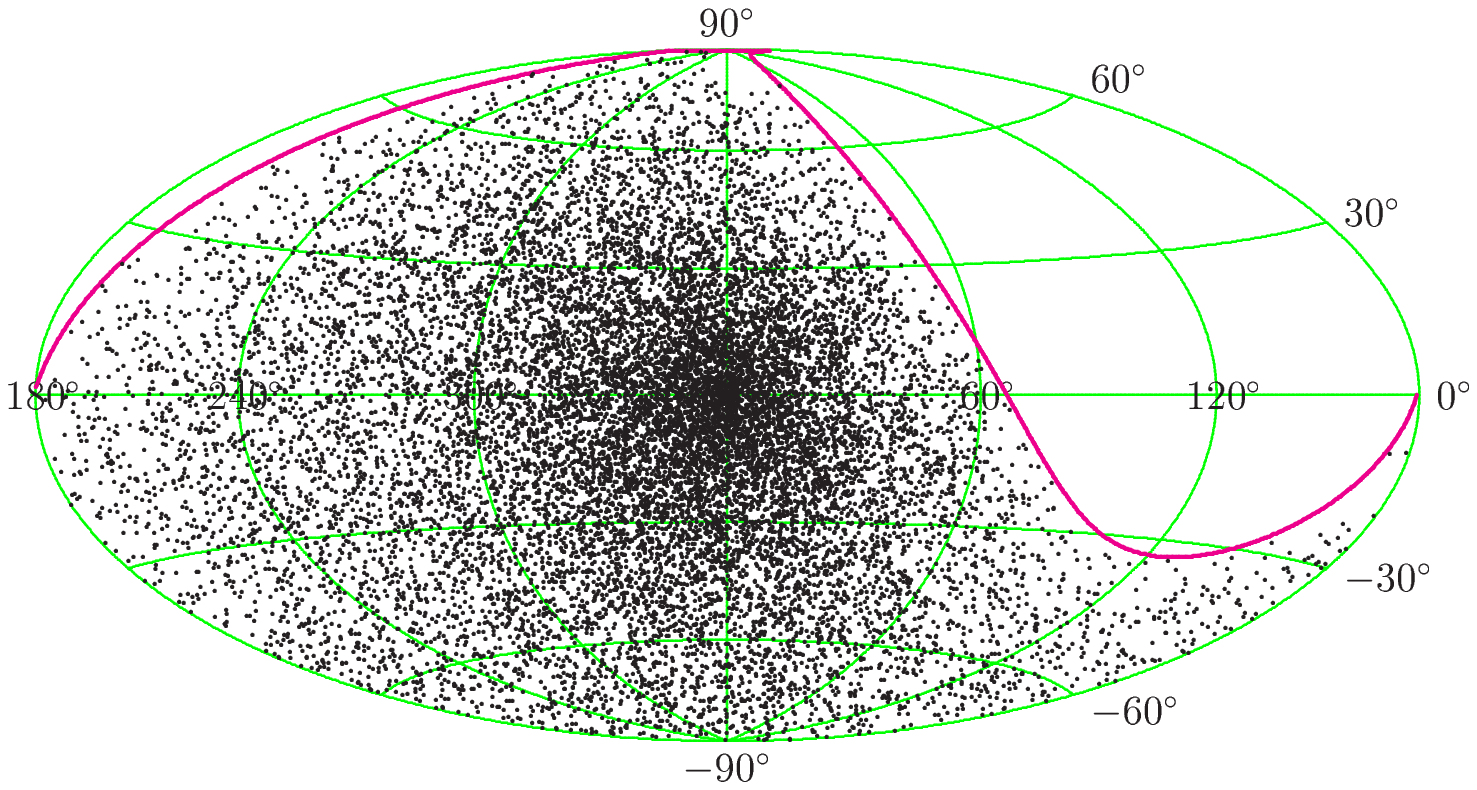}
\end{center}
\caption{(a) The sky map of arrival directions of 316 SUGAR events
with zenith angle $z\le60^\circ$ and energy $E\ge1.0\times10^{19}$ eV.
(b) and (c) show $316\times50$ simulated events for the isothermal
halo model with $R_c=4$ kpc and the NFW halo model with $R_s=15$ kpc,
respectively.  Galactic coordinates are used; the solid line is the
boundary of the SUGAR acceptance with the zenith angle cut of
$60^\circ$.}
\end{figure}

\section{Kolmogorov-Smirnov test}

The most direct way to compare the arrival direction distribution of
the real data and that predicted by the superheavy dark matter model
is to use the two-dimensional K-S test over the pair $(\alpha,\delta)$
or $(l,b)$. Here, we use a different method which is simpler and gives
better results. Since the cosmic rays in the halo model are
distributed symmetrically around the Galactic center, we analyze the
distribution of cosmic rays over the angle $\theta$ between their
arrival directions and the direction to the Galactic center. We use
the one-dimensional K-S test to compare the SUGAR data and model
predictions. Since one-dimensional K-S test is
reparametrization-invariant, we use $x=\cos\theta$ which is more
convenient in computation.

The K-S test makes use of the cumulative probability distribution function,
which is defined by
\begin{equation}
S_N(x)=\frac{1}{N}\sum_i\theta(x_i-x),
\end{equation}
where $N$ is the number of data points. For the distribution of
arrival directions expected in the Galactic halo model, we generate
large number $M$ of simulated events and obtain corresponding
cumulative distribution $S_{M}(x)$. We then calculate the K-S
statistic
\begin{equation}
D=\max\left|S_{N}(x)-S_{M}(x)\right|,
\end{equation}
which determines the probability that the two sets of events are drawn
from the same distribution \cite{nrc}. Small probabilities indicate
that the two sets significantly differ.  Figure~2 shows the cumulative
distribution functions used in the K-S test. The difference between the
SUGAR data and halo models is clearly seen (it can be expected already
from Figure~1).

%
%
\begin{figure}
\begin{center}
\includegraphics[width=0.6\textwidth]{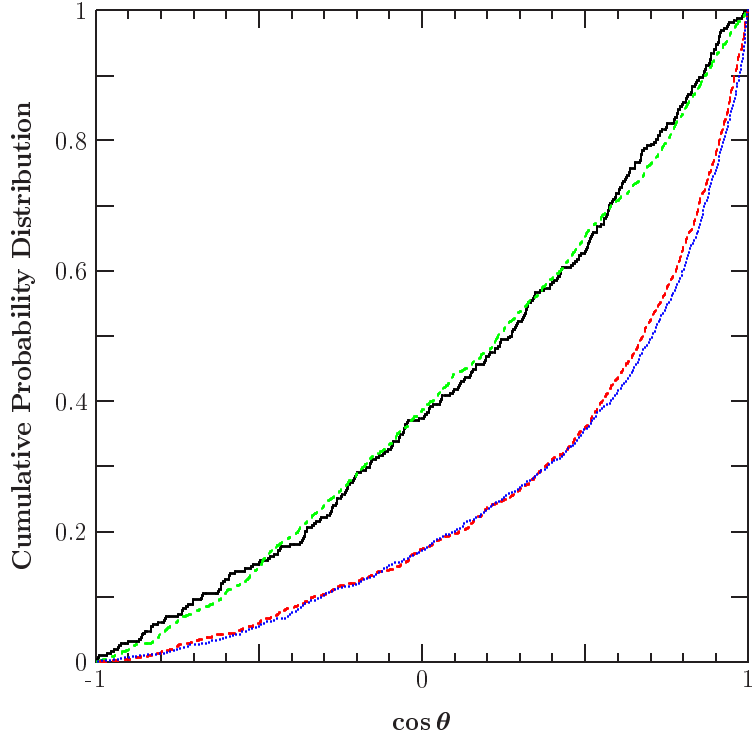}
\end{center}
\caption{Cumulative probability distribution functions
used in Kolmogorov-Smirnov test
of 316 SUGAR data ($z\le60^\circ$ and $E\ge1\times10^{19}$ eV, solid),
$4\times316$ simulated data of isotropic distribution (dash-dotted),
the isothermal halo model ($R_c=4$ kpc, dashed),
and the NFW halo model ($R_s=15$ kpc, dotted).}
\end{figure}

As mentioned in the Introduction, the SHDM model does not explain
clustering and correlations of UHECR with BL Lacs, so another
(extragalactic) component is needed which is isotropic at large
angles. The SHDM model itself contains the isotropic extragalactic
contribution coming from SHDM decays in other galaxies.
Thus, let us consider a generic model where UHECR flux consists of
the isotropic and Galactic halo components (their relative strength
in the SHDM models is discussed in the next section).
An important quantity is the 
fraction $\eta$ of cosmic rays coming from the Galactic halo,
\begin{equation}
\eta(E_{\rm min}) =
\frac{\int_{E_{\rm min}}f_{\rm G}(E)\,dE}%
{\int_{E_{\rm min}}\left(f_{\rm G}(E)+f_{\rm U}(E)\right)\,dE},
\end{equation}
where $f_{\rm G}(E)$ and $f_{\rm U}(E)$ are the energy spectra of
cosmic rays from the Galactic halo and from the uniform background,
respectively.  We consider $\eta$ as a function of the lower energy
cut $E_{\rm min}$.  When two spectra $f_{\rm G}(E)$ and $f_{\rm U}(E)$
have quite different shape, $\eta(E_{\rm min})$ depends sensitively on
the value of $E_{\rm min}$.  Instead of following a specific model to
evaluate $\eta(E_{\rm min})$, we use $\eta$ as a parameter and
calculate the probability to get the SUGAR data distribution. In this
way bounds on $\eta$ at a given energy cut can be obtained.

Figure~3 shows the dependence of the probability that all cosmic rays
come from the Galactic halo ($\eta=1$) as a function of the energy cut
$E_{\rm min}$ and the halo parameters $R_c$ and $R_s$. The dependence
on $R_s$ in the case of NFW halo profile is rather weak. At realistic
values of parameters $R_c\sim4$ kpc and $R_s\sim15$ kpc the two halo
models give similar constraints. Dominant fraction of halo component
is excluded up to energies as high as $5\times 10^{19}$~eV. For higher
energies, the constraints become progressively weaker; the reason may
be much smaller statistics.
%
%
\begin{figure}
\begin{center}
\includegraphics[width=0.45\textwidth]{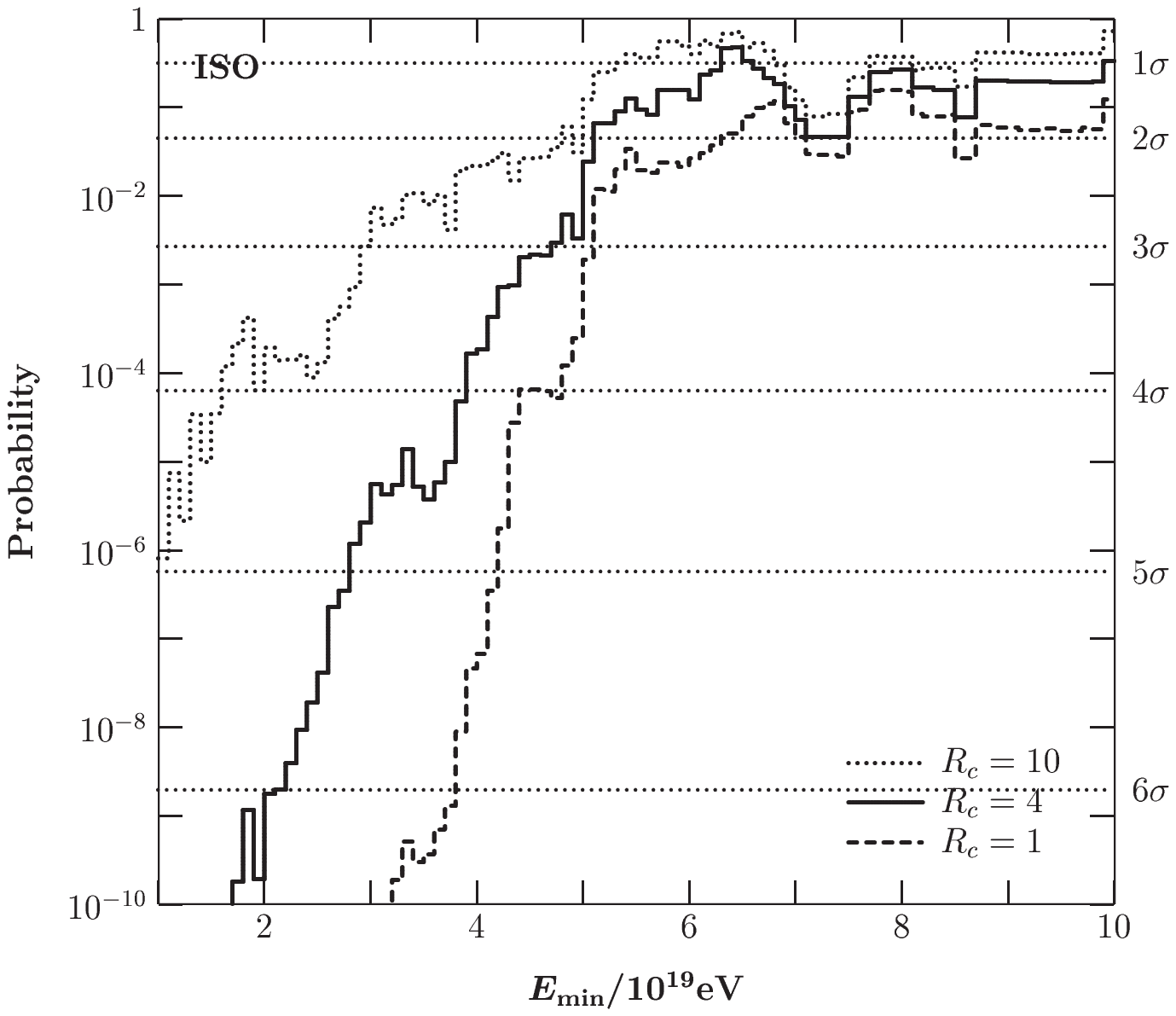}
\includegraphics[width=0.45\textwidth]{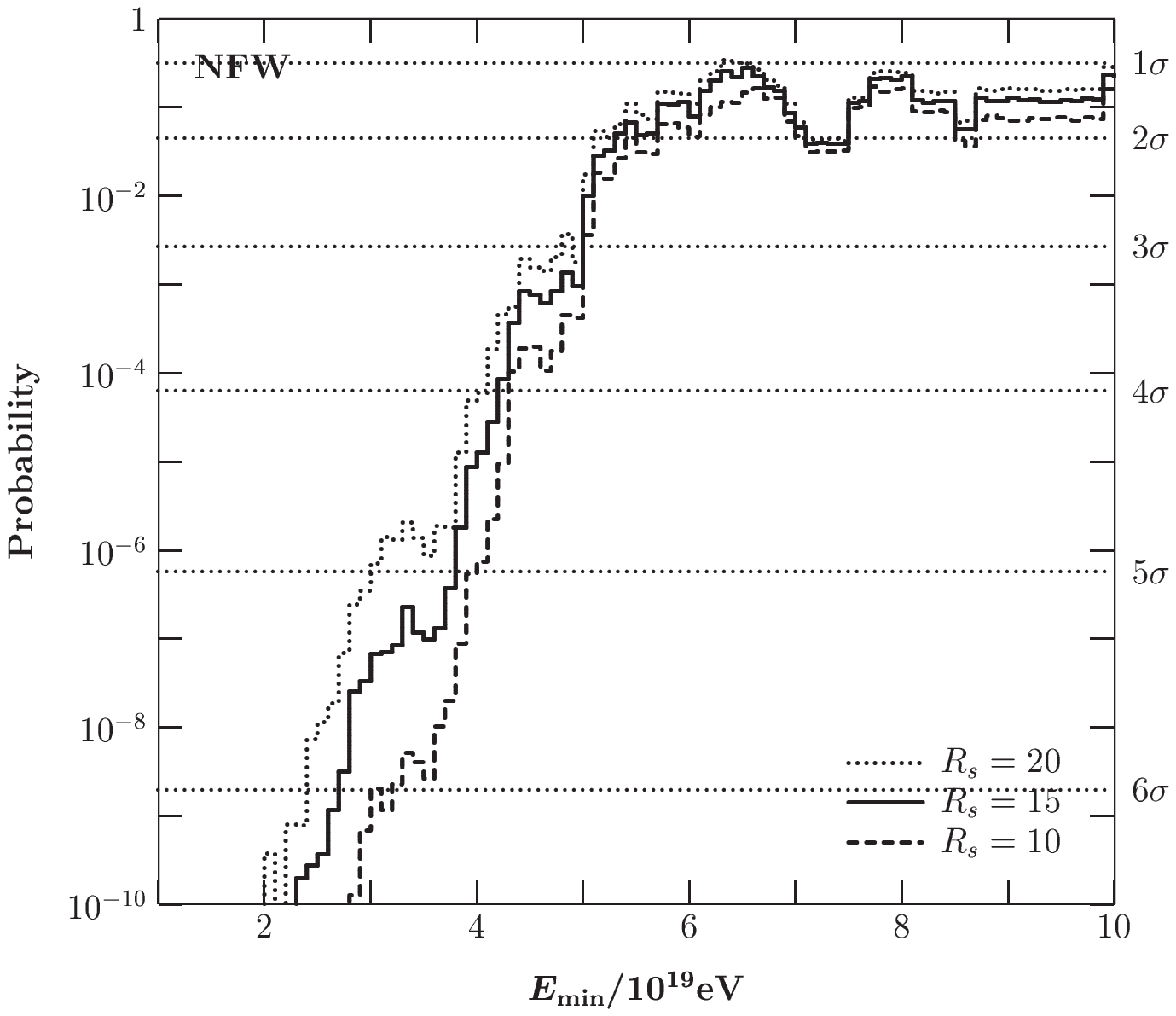}
\includegraphics[width=0.45\textwidth]{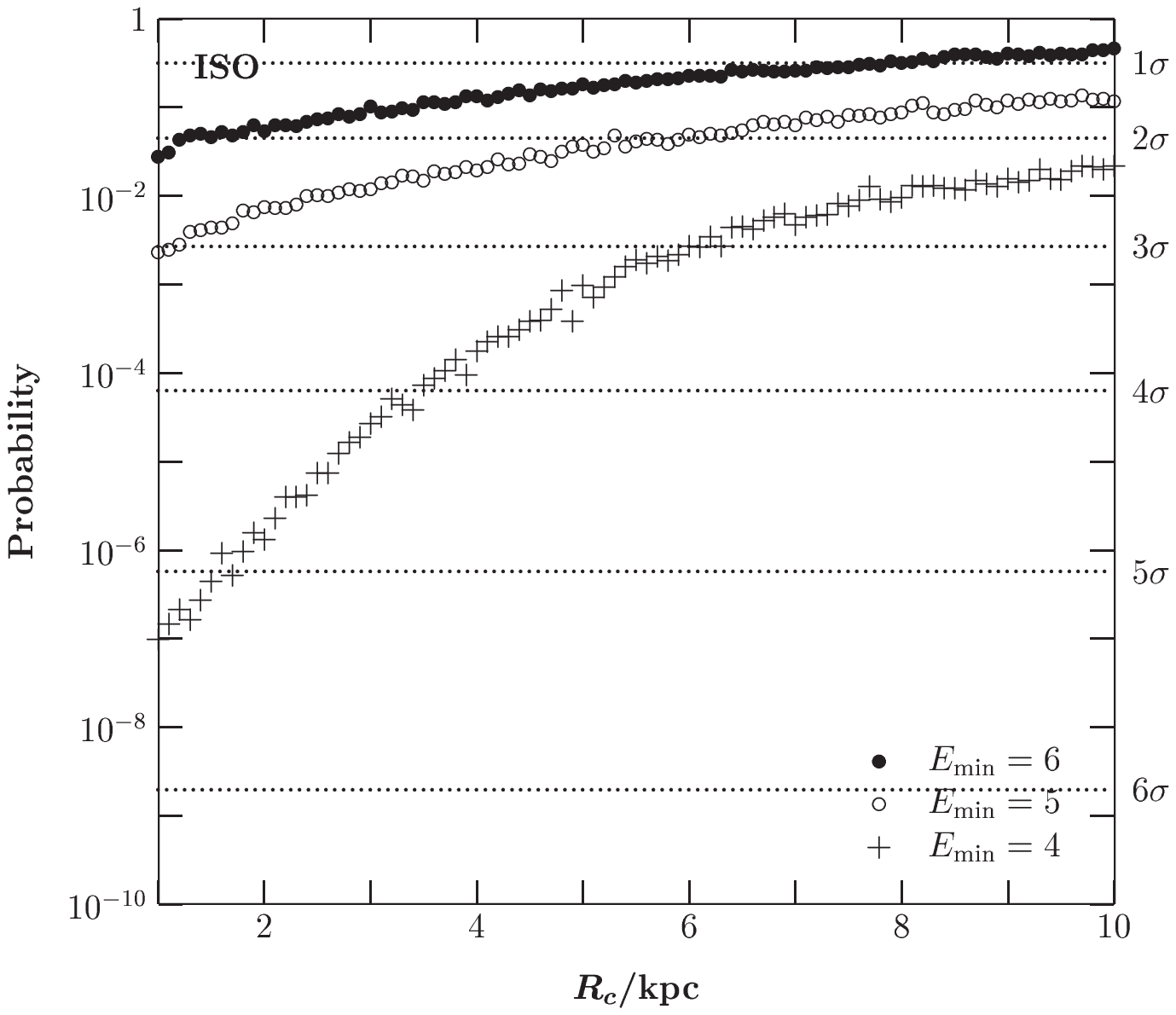}
\includegraphics[width=0.45\textwidth]{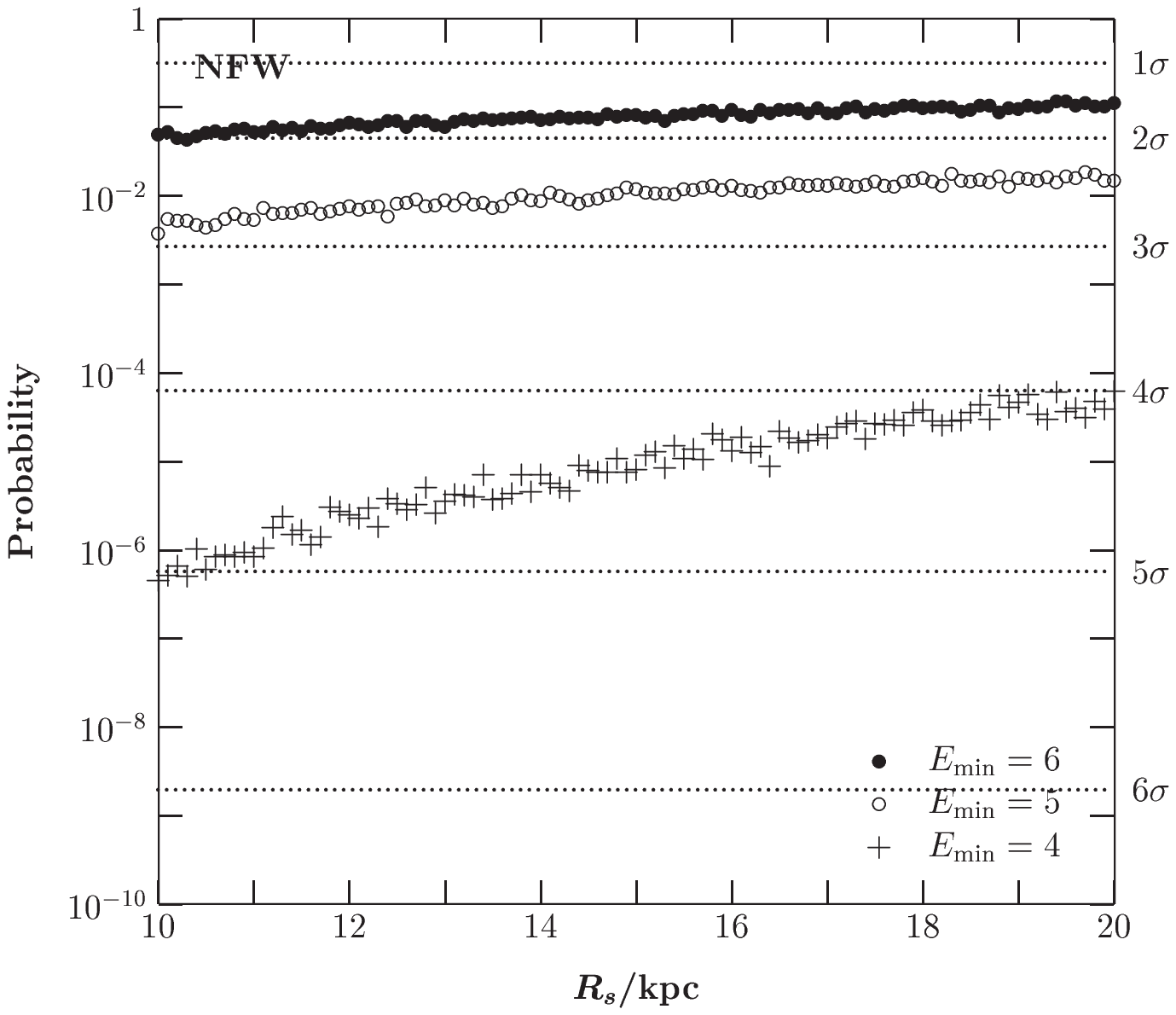}
\end{center}
\caption{Probability that all cosmic rays come from the Galactic halo
($\eta=1$).  Up: dependence on energy cut for different values of
$R_c$ (isothermal halo) and $R_s$ (NFW halo).  Down: dependence on
halo parameter $R_{c,s}$ for different values of energy cut.}
\end{figure}

Figure~4 shows the lines of constant probability (corresponding to
$1\sigma$ -- $4\sigma$ levels) in the plane of the model parameters
$R_c,R_s$ and the fraction of halo component $\eta$. Two energy cuts
are represented: $E_{\rm min} = 1\times10^{19}$~eV (upper panels) and
$E_{\rm min} = 4\times10^{19}$~eV (lower panels). In the case of NFW
halo profile, isotropy of SUGAR data at $E>1\times10^{19}$~eV excludes
fractions of halo component larger than 15\% at $2\sigma$ level, while
for $E>4\times10^{19}$~eV the halo fraction has to be less than 45\%.
Constraints are similar for isothermal halo profile eq.(\ref{eq-iso}) at
$R_c\sim 4$~kpc. 
%
%
\begin{figure}
\begin{center}
\includegraphics[width=0.45\textwidth]{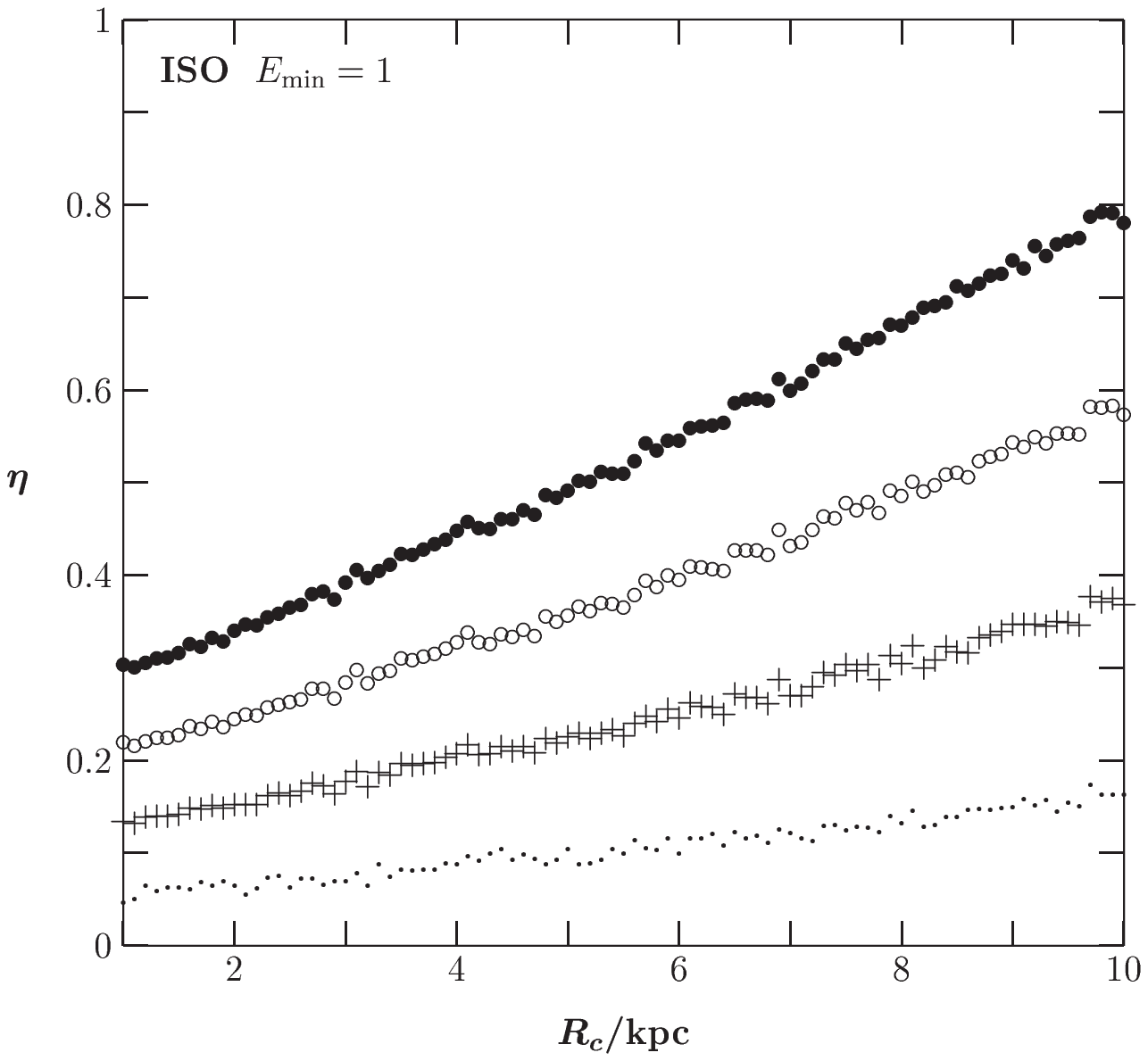}
\includegraphics[width=0.45\textwidth]{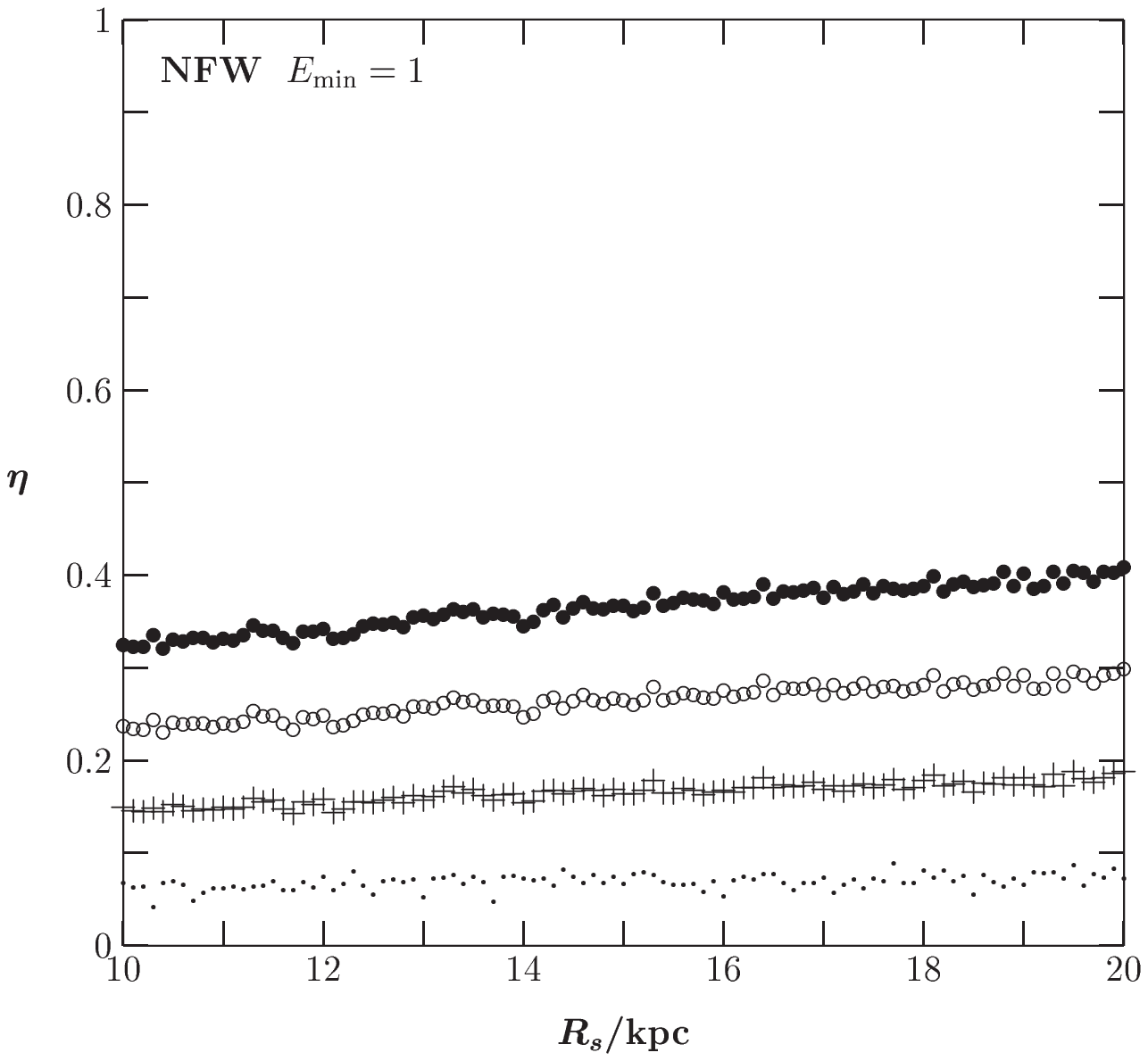}
\includegraphics[width=0.45\textwidth]{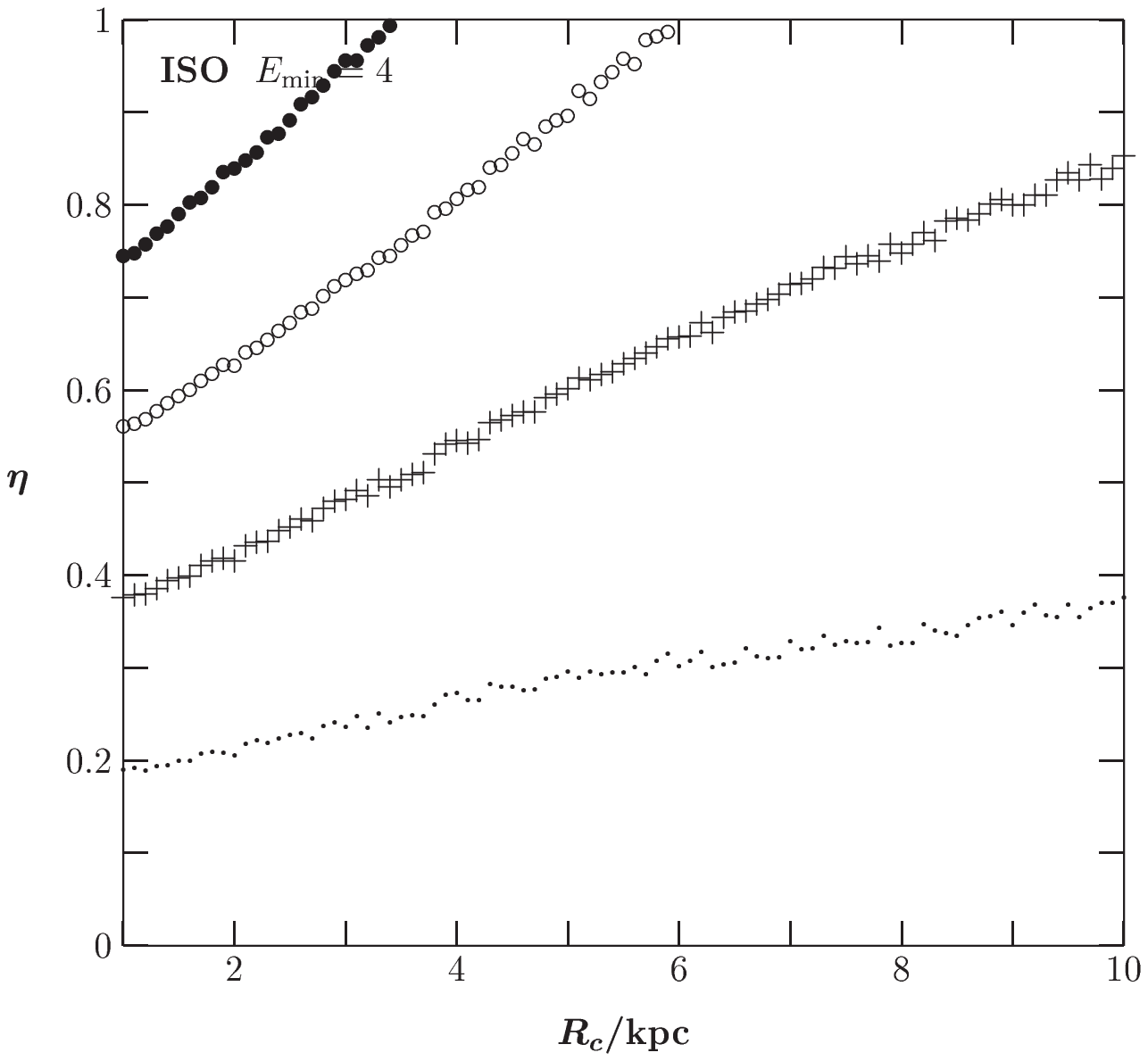}
\includegraphics[width=0.45\textwidth]{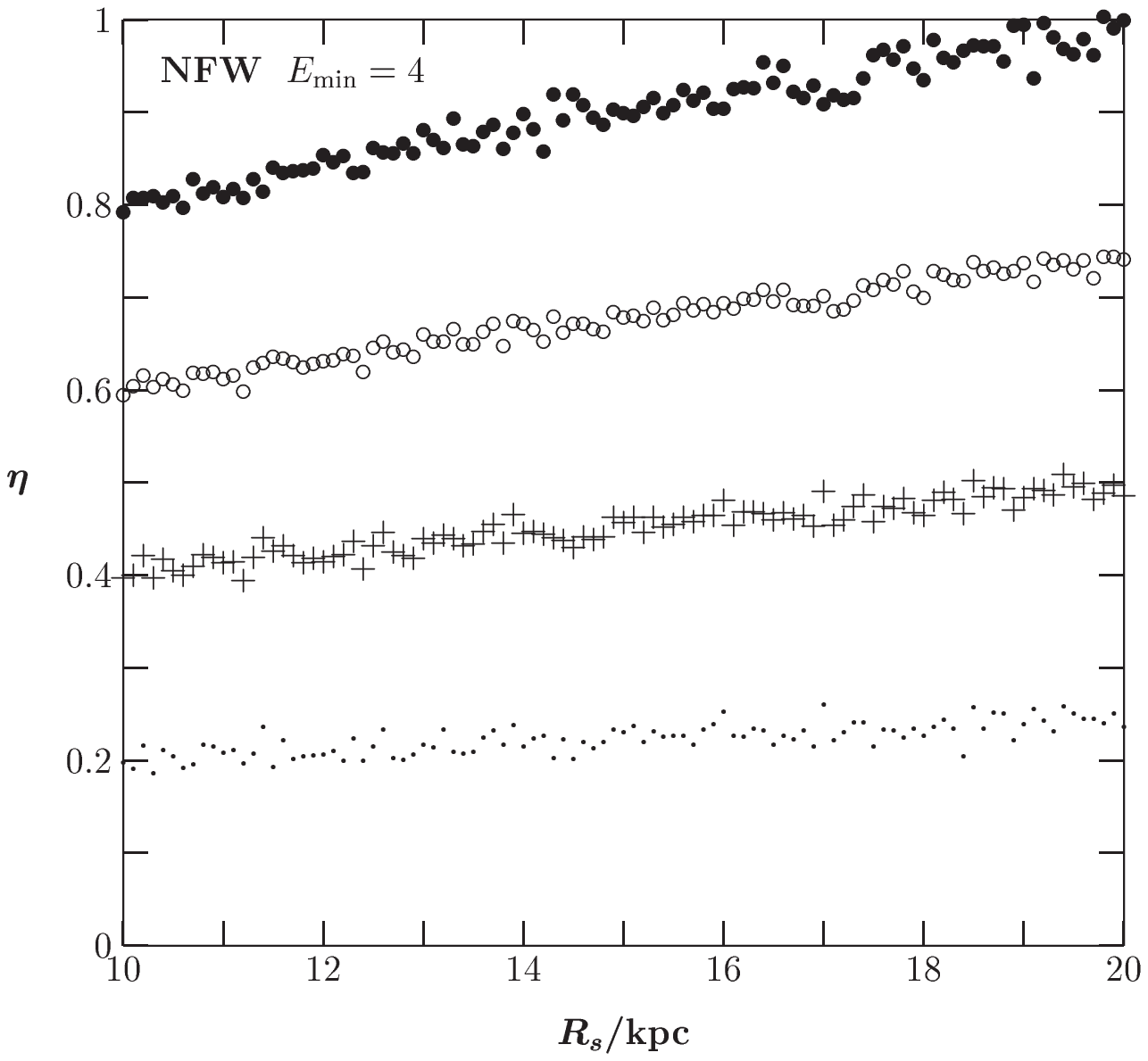}
\end{center}
\caption{Exclusion plots (lines of constant probability) for the
two-component model described in the text on the $\eta$ -- $R_c$
(isothermal halo, left panels) and $\eta$ -- $R_s$ (NFW halo, right
panels) plane. Two energy cuts are presented, $E_{\rm
min}=1\times10^{19}$~eV (upper panels) and $E_{\rm
min}=4\times10^{19}$~eV (lower panels).
Symbols $\cdot$, $+$, $\circ$ and $\bullet$ represent $1\sigma$ --
$4\sigma$ lines, respectively.}
\end{figure}

Finally, let us comment on how robust are the results presented here.
One thing one should check is the influence of the Galactic magnetic
field during the propagation. This can be important because
significant portion of cosmic rays from the Galactic halo are produced
near the Galactic center and experience strong magnetic fields. To see
the effect of this spread on our results, we added a Gaussian random
deflections with mean deviation up to $10^\circ$ in the simulated data
set. This effect also mimics possible poor angular resolution of
SUGAR events. We found that results presented above are stable against
this modification. The maximum change occurs
at $E_{\rm min}\approx1\times10^{19}$ eV and $\eta\approx1$.
For the moderate energy cut or with the uniform background component added,
its effect becomes less than 10\% and can be completely neglected.

\section{Application to SHDM model}

In order to see what the above constraints imply for a given SHDM model,
one has to estimate the strength of the halo (anisotropic)
component relative to the extragalactic (isotropic) one.
The latter consists of a part related to SHDM decays outside of the Galaxy
and a part due to astrophysical sources. Let us estimate the former. 
The Galactic and extragalactic contributions in the SHDM model can be
written as 
\[
 F_G(E) = C  \int d^3x {n(x)\over (\vec R - \vec x)^2},
\]
\[
F_{\rm ext}(E) = C \int d^3x {\bar n \over r^2} w(r). 
\] 
Here $\bar n$ and $n(x)$ are the average and halo dark matter
densities, $\vec R$ is the vector from Galactic center to the Earth,
while $w(r)$ is the weight function which accounts for the flux
attenuation at large distances. The proportionality coefficient $C$
which accounts for the efficiency of UHECR production from SHDM is the
same in both cases. It therefore cancels in the ratio of the two
contributions.

The extragalactic part can be written in the form 
\[
F_{\rm ext} =  2\pi C \bar n  R_{\rm eff},  
\]
where by definition $R_{\rm eff}= \int dr w(r)$. This quantity
represents the size of the region from which the extragalactic flux is
collected (for instance, for very high-energy protons $R_{\rm eff}$ is
the GZK distance). It can be estimated as the attenuation length
divided by the spectral index. The factor $2\pi$ in this equation
takes into account the fact that SUGAR sees roughly half of the sky.

To estimate the halo contribution we take the NFW profile at
$R_s=15$~kpc and calculate the flux coming from half of a sphere
overlooking the Galactic center. The result can be written as follows,
\[
F_G \simeq  2\pi C \bar n \times 5000\mbox{~Mpc}.
\]
When obtaining this number the ratio of the local to average dark
matter density was taken to be $2\times10^{5}$. This number is
uncertain by a factor 2 or so.

The ratio of the two components is 
\[
{F_{\rm ext} \over F_G } \simeq {R_{\rm eff} \over 5000\mbox{~Mpc}}.
\]
The UHECR flux in the SHDM models is dominated by photons 
which have attenuation length less than 100~Mpc in the range of
energies $10^{19}-10^{20}$~eV even at very weak assumptions about
radio photon background \cite{BhattSigl}. We therefore arrive at the
conclusion that in the SHDM models the halo component dominates
already at energies around $10^{19}$~eV, i.e., quite
below the GZK cutoff.  Since the halo component dominates, the bounds
on the fraction $\eta$ in Figures~3 and 4
essentially apply to the total fraction of UHECR which originate from
decays of SHDM particles. 

The situation becomes even better at high energies where the
extragalactic component dies away and the halo contribution reaches
100\%. In this energy range $\sim 60$ uniformly distributed events
would be enough to rule out the SHDM model at $4-5\sigma$ level
following the technique developed here.

\section{Conclusions}

We have used the SUGAR data to check the viability of the hypothesis
that a part of the highest-energy cosmic rays originate from decays of
relic SHDM particles. We found that the absence of the anisotropy
toward the Galactic center in the SUGAR data imposes strong
constraints on corresponding fraction of UHECR. At 95\% confidence
level, the fraction of SHDM-related cosmic rays should be less than
20\% and 50\% at $E>1\times10^{19}$~eV and $E>4\times10^{19}$~eV,
respectively. This implies that the SHDM component of UHECR, if
exists, must have harder spectrum than the uniform one and 
should take over at energies higher than $4\times10^{19}$~eV.  

\acknowledgments The authors are grateful to S.~Dubovsky for helpful
discussions. This work is supported by the Swiss Science Foundation,
grant 20-67958.02.

\bigskip\noindent
{\bf Note Added}: While completing this work, we received a
preprint by M. Kachelrie\ss\ and D. V. Semikoz \cite{ks03} where
similar results were obtained. We thank M. Kachelrie\ss\ and
D. V. Semikoz for sending us the preprint before publication.

\def\arxiv#1{arXiv:#1}
\def\journal#1#2#3#4{#1 #2, #3 (#4)}
\def\aap{Astron.\ Astrophys.\ }
\def\apj{Astrophys.\ J.\ }
\def\app{Astropart.\ Phys.\ }
\def\araa{Annu.\ Rev.\ Astron.\ Astrophys.\ }
\def\cap{Comments Astrophys.\ }
\def\ijmp{Int.\ J.\ Mod.\ Phys.\ }
\def\jhep{JHEP\ }
\def\jpg{J.\ Phys.\ G }
\def\jetp{JETP\ }
\def\jetpl{JETP Lett.\ }
\def\mnras{MNRAS\ }
\def\mpl{Mod.\ Phys.\ Lett.\ }
\def\na{New\ Astronomy\ }
\def\nature{Nature\ }
\def\np{Nucl.\ Phys.\ }
\def\pan{Phys.\ Atom.\ Nucl.\ }
\def\pl{Phys.\ Lett.\ }
\def\pr{Phys.\ Rev.\ }
\def\prd{Phys.\ Rev.\ D }
\def\prl{Phys.\ Rev.\ Lett.\ }
\def\prt{Phys.\ Rep.\ }
\def\pzetf{Pisma Zh.\ Exp.\ Teor.\ Fiz.\ }
\def\pz{Phys.\ Z.\ }
\def\rmp{Rev.\ Mod.\ Phys.\ }
\def\rmp{Rev.\ Mod.\ Phys.\ }
\def\rpp{Rep.\ Prog.\ Phys.\ }
\def\spu{Sov.\ Phys.\ Usp.\ }
\def\ssr{Space Science Reviews }
\def\yf{Yad.\ Fiz.\ }
\def\etal{{\it et al.}}


\begin{thebibliography}{99}

\bibitem{gzk}
K. Greisen, \journal{\prl}{16}{748}{1966};
G. T. Zatsepin and V. A. Kuzmin,
\journal{\jetpl}{4}{78}{1966} [\journal{\pzetf}{4}{114}{1966}].

\bibitem{agasa}
M. Takeda \etal, \journal{\prl}{81}{1163}{1998};
N. Hayashida \etal, \journal{\apj}{522}{225}{1999}.

\bibitem{hires}
T. Abu-Zayyad \etal, \arxiv{astro-ph/0208301}.

\bibitem{cluster}
M. Takeda \etal, \arxiv{astro-ph/9902239};
Y. Uchihori \etal, \journal{\app}{13}{151}{2000}.

\bibitem{bllacs}
P.~G.~Tinyakov and I.~I.~Tkachev,
JETP Lett.\  {\bf 74}, 445 (2001)
[Pisma Zh.\ Eksp.\ Teor.\ Fiz.\  {\bf 74}, 499 (2001)]
[arXiv:astro-ph/0102476].
P.~G.~Tinyakov and I.~I.~Tkachev,
Astropart.\ Phys.\  {\bf 18}, 165 (2002)
[arXiv:astro-ph/0111305].
D.~S.~Gorbunov, P.~G.~Tinyakov, I.~I.~Tkachev and S.~V.~Troitsky,
Astrophys.\ J.\  {\bf 577}, L93 (2002)
[arXiv:astro-ph/0204360].

\bibitem{Kachelriess:2003yy}
M.~Kachelriess, D.~V.~Semikoz and M.~A.~Tortola,
arXiv:hep-ph/0302161.

\bibitem{Berezinsky:2002vt}
V.~Berezinsky, A.~Z.~Gazizov and S.~I.~Grigorieva,
arXiv:astro-ph/0210095.

\bibitem{shdm}
V. Berezinsky, M. Kachelriei\ss\ and A. Vilenkin,
\journal{\prl}{79}{4302}{1997};
V. A. Kuzmin and V. A. Rubakov,
\journal{\pan}{61}{1028}{1998} [\journal{\yf}{61}{1122}{1998}]

\bibitem{bdk02}
P. Blasi, R. Dick and E. W. Kolb, \journal{\app}{18}{57}{2002}.

\bibitem{kt99}
V. A. Kuzmin and I. I. Tkachev, \journal{\prd}{59}{123006}{1999};
D. J. H. Chung, E. W. Kolb and A. Riotto, \journal{\prd}{60}{063504}{1999}.

\bibitem{dt98}
S. L. Dubovsky and P. G. Tinyakov, \journal{\jetpl}{68}{107}{1998};
V. Berezinsky, P. Blasi and A. Vilenkin, \journal{\prd}{58}{103515}{1998}.
G.~A.~Medina Tanco and A.~A.~Watson,
Astropart.\ Phys.\  {\bf 12}, 25 (1999)
[arXiv:astro-ph/9903182].

\bibitem{Hayashida:1999ab}
N.~Hayashida {\it et al.}  [AGASA Collaboration],
arXiv:astro-ph/9906056.

\bibitem{bsw99}
A. Benson, A. Smialkowski and A. W. Wolfendale, \journal{\app}{10}{313}{1999}.

\bibitem{sugar}
M. M. Winn \etal, \journal{\jpg}{12}{653}{1986};
The complete catalogue of SUGAR data in ``Catalogue of highest energy cosmic rays No.~2'', ed. WDC-C2 for Cosmic Rays (1986).

\bibitem{iso}
J. Caldwell and J. Ostriker, \journal{\apj}{251}{61}{1981};
J. Bahcall et al., \journal{\apj}{265}{730}{1983}.

\bibitem{nfw96}
J. F. Navarro, C. S. Frenk and S. D. M. White, \journal{\apj}{462}{563}{1996}.

\bibitem{nrc}
W. H. Press, S. A. Teukolsky, W. T. Vetterling and B. P. Flannery,
Numerical Recipes in C, Cambridge University Press (1992, Cambridge).

\bibitem{BhattSigl} 
P.~Bhattacharjee and G.~Sigl,
Phys.\ Rept.\  {\bf 327}, 109 (2000)
[arXiv:astro-ph/9811011].

\bibitem{ks03}
M. Kachelrie\ss\ and D. V. Semikoz, \arxiv{astro-ph/0306282}

\end{thebibliography}
\end{document}